# Automatic prediction of cognitive and functional decline can significantly decrease the number of subjects required for clinical trials in early Alzheimer's disease

Neda SHAFIEE[1], Mahsa Dadar[2], Ph.D., Simon Ducharme[1,3], M.D. M.Sc., D. Louis Collins[1], Ph.D.
for the Alzheimer's Disease Neuroimaging Initiative[*]

[1] McConnell Brain Imaging Centre, Montreal Neurological Institute, McGill University, Montreal, Quebec (QC), Canada.

[2] CERVO Brain Research Center, Centre intégré universitaire santé et services sociaux de la Capitale Nationale, Québec, QC

[3] Douglas Mental Health University Institute, Department of Psychiatry, McGill University, Montreal, Quebec (QC), Canada.

**Corresponding Author Information:**

Neda Shafiee, McConnell Brain Imaging Centre, Montreal Neurological Institute, 3801 University Street, Room WB320, Montréal, QC, H3A 2B4

Email : neda.shafiee@mail.mcgill.ca

---

[*] Data used in preparation of this article were obtained from the Alzheimer's Disease Neuroimaging Initiative (ADNI) database (adni.loni.usc.edu). As such, the investigators within the ADNI contributed to the design and implementation of ADNI and/or provided data but did not participate in analysis or writing of this report. A complete listing of ADNI investigators can be found at:
https://adni.loni.usc.edu/wp-content/uploads/how_to_apply/ADNI_Acknowledgement_List.pdf

## ABSTRACT

**INTRODUCTION:** Heterogeneity in the progression of Alzheimer's disease makes it challenging to predict the rate of cognitive and functional decline for individual patients. Tools for short-term prediction could help enrich clinical trial designs and focus prevention strategies on the most at-risk patients.

**METHOD:** We built a prognostic model using baseline cognitive scores and MRI-based features to determine which subjects with mild cognitive impairment remained stable and which functionally declined (measured by a two-point increase in CDR-SB) over 2 and 3-year follow-up periods, periods typical of the length of clinical trials.

**RESULTS:** Combining both sets of features yields 77% accuracy (81% sensitivity and 75% specificity) to predict cognitive decline at 2 years (74% accuracy at 3 years with 75% sensitivity and 73% specificity). Using this tool to select trial participants yields a 3.8-fold decrease in the required sample size for a 2-year study (2.8-fold decrease for a 3-year study) for a hypothesized 25% treatment effect to reduce cognitive decline.

**DISCUSSION:** This cohort enrichment tool could accelerate treatment development by increasing power in clinical trials.

# 1. Introduction:

Alzheimer's disease (AD) is a neurodegenerative disorder characterized by abnormal accumulation of amyloid-β (Aβ) and intracellular neurofibrillary tangles (NFTs) in the brain resulting in progressive synaptic dysfunction, irreversible neuronal loss, and cognitive deficits [1], [2]. This pathological process gradually develops over many years, with a long asymptomatic phase before a clinical diagnosis of AD [3]. Patients in the early stages of AD dementia experience decline in their cognitive abilities at different rates, with some patient progressing very fast while a large portion of patients remain stable [4], [5]. This heterogeneity increases the complexity of treatment development. After numerous failures of candidate drugs for dementia due to AD, the field has moved toward clinical trials at an earlier stage (mild cognitive impairment with proven AD biomarkers [6], [7]). However, those trials generally do not factor the marked inter-individual differences in rates of progression in subjects with MCI, which can have a profound effect on the outcome of trials [8], [9]. Consequently, there is still no approved disease modifying treatment for MCI due to AD. Accurately predicting the progression rate in individual patients with mild cognitive impairment and mild dementia due to AD would enable the enrichment of patient populations in clinical trials by increasing the mean cognitive/functional decline over the trial duration, and therefore facilitating the demonstration of the treatment effect (or the absence of treatment effect). This in turn could lead to potentially faster, more efficient candidate drug testing.

In order the be generalizable to the population after drug approval, tools to predict future progression in MCI would have to be based on readily available measures in clinical practice, such as brain MRI and cognitive tests. Indeed, AD is associated with a stereotypical pattern of early cerebral atrophy in the medial temporal lobe (MTL) limbic regions including entorhinal cortex (EC) and hippocampus (HC) [1]. The early degeneration in MTL limbic structures consistent with early memory deficits provides the anatomical basis to use MRI-based measures of atrophy as valid markers of disease state and progression [10], [11].

We have previously developed Scoring by Nonlocal Image Patch Estimator (SNIPE) as a grading metric to measure AD-related structural alterations in brain anatomy, with applications to both hippocampal and entorhinal structures [12]. Based on this nonlocal patch-based framework, SNIPE estimates the structural similarity of a new subject under study to a number of templates present in a training library consisting of cognitively normal subjects and patients with AD. In our previous work, we showed that baseline SNIPE scores could differentiate patients with mild cognitive impairment (MCI) that remain stable vs those that progress to AD [13], and that baseline SNIPE scores enable AD prediction in a group of cognitively intact subjects seven years before the clinical diagnosis of AD dementia [14]. More recently, we demonstrated that combining MRI features and neurocognitive test results at baseline could yield 78% accuracy in prediction of conversion from MCI to AD at 2 and 3 years before diagnosis of AD (and up to 87% accuracy, five years before diagnosis) [15].

While these results were promising, conversion to AD as a categoricaldiagnosis may be too late an event when testing new neuroprotective therapies. In this study, we investigated the ability of our models to predict cognitive and functional decline (as opposed to categorical change in diagnosis from MCI to mild dementia) in a cohort of patients with mild AD similar to those chosen for recent clinical trials [16], [17]. Using only baseline cognitive test results and baseline MR-driven features, we evaluate the accuracy, sensitivity, and specificity of our model to predict decline over two- and three-year follow-up periods, durations commonly used in clinical trials. Cognitive decline is defined as an increase in global CDR-SB (CDR-Sum of boxes) score [18]. Finally, we evaluate the potential use of our proposed technique as a screening tool for enrichment in clinical trials targeting patients likely to experience cognitive decline in near future.

## 2. Methods

### 2.1. Dataset

Data used in this study include patients with mild Alzheimer's disease from the Alzheimer's Disease Neuroimaging Initiative (ADNI) database (adni.loni.usc.edu). The ADNI was launched in 2003 as a public-private partnership led by Principal Investigator Michael W. Weiner, MD. The primary goal of ADNI has been to test whether serial MRI, positron emission tomography (PET), other biological markers, and clinical and neuropsychological assessment can be combined to measure the progression of MCI and early AD.

In this work, we selected subjects with mild AD from ADNI study for which T1 MRI data and MoCA (Montreal Cognitive Assessment) scores were available at baseline. The inclusion of MoCA limited this study to ADNI2 and ADNI-GO datasets, since this measurement was not included in ADNI1 dataset. The key inclusion criteria here are similar to those used for current clinical trials in MCI cohorts: 1) A Clinical Dementia Rating (CDR)-Global Score of 0.5, 2) An MMSE score between 24 and 30 (inclusive), and 3) having a positive amyloid Positron Emission Tomography (PET) scan or CSF amyloid positivity. Application of these criteria reduced the number of subjects available at baseline in ADNI2 and ADNI-GO to 312. These subjects were labeled as either stable or progressive based on a 2-point increase in their global CDR-SB score from a total possible of 18 points [18]. Here, we refer to the stable and progressive mild AD subjects as pMCI and sMCI, respectively.

**Table 1**: Dataset Information

|  | 2 years follow-up | 3 years follow-up |
| --- | --- | --- |
| **pMCI** | 55 | 63 |
| **sMCI** | 155 | 108 |
| **pMCI:sMCI ratio** | 0.355 | 0.583 |
| **Age at baseline** | 72.5±6.7 | 71.9±6.6 |
| **% Male** | 54.3 | 55.6 |

### 2.2. Preprocessing

All the selected T1 MR images were preprocessed using a fully automatic pipeline. This pipeline includes denoising [19], correction of intensity inhomogeneity using N3 [20], and intensity normalization. MRI scans were then registered to pseudo-Talairach stereotaxic space [21], [22] using a population-specific template [23]. Brain extraction was then performed using BEaST [24].

*2.3. MRI features: SNIPE scoring*

To automatically segment HC and EC, a multi-template non-local patch-based method has been used [25]. This method uses a set of MRI volumes with manually segmented HC and EC as training library. The target patch is then weighted based on how much it resembles each patch in the training dataset. The final label of the patch (targeted structure or background) was assigned based on a weighted average of all similar patches.

The SNIPE grading or scoring of the HC and EC is then achieved by estimating the patch similarity of the subject under study to different training populations: normal controls and patients with Alzheimer's dementia [12], [13]. Following the same linear regression method used in [26], SNIPE scores are corrected for age and sex based on the normal control population. Visual quality control was performed on all processed MR datasets.

*2.4. Classification*

Our feature set contains age, sex, cognitive test scores including Alzheimer's Disease Assessment Scale (ADAS), MoCA, Rey Auditory Verbal Learning Task (RAVLT), Mini-Mental State Examination (MMSE), and MR-based z-scored features (SNIPE scores for HC and EC) from baseline data that are used as input to the classifier.

Since the number of sMCI and pMCI subjects were not the same, and standard methods may have difficulty with such imbalanced data, we used a balanced random forest algorithm to train our predictive model [27]. This method down-samples the majority class and trains the trees of the random forest based on a more balanced data set.

We trained our prognostic model using different combinations of features drawn from baseline visits. These classifiers were trained either using MRI-driven SNIPE scores and age, neurocognitive scores and age, or a combination of both SNIPE and neurocognitive scores plus age, and each model was validated using 10-fold cross-validation. The classification performance for both follow-up periods (i.e., 2 and 3 years) was evaluated based on the measured sensitivity, specificity, and accuracy.

*2.5. Power analysis*

Following the method used in [28], we estimated the required sample size to detect a reduction in the mean annual rate of cognitive decline based on CDR-SB score. This method assumes that rates of decline are linear for each subject. We used a two-sided test and set the standard significance level to 0.05 with a power of 80%. The required sample size per arm was estimated using the following formula (Fitzmaurice et al. 2011):

$$n = \frac{2(\sigma_s^2 + \frac{\sigma_\epsilon^2}{\sum(t_j - \bar{t})^2})(Z_{1-\frac{\alpha}{2}} + Z_{1-\beta})}{\Delta^2} \tag{1}$$

Where α and 1-β are the significance level and power and $\bar{t}$ represents the mean measure time. $\sigma_s^2$ and $\frac{\sigma_\epsilon^2}{\sum(t_j-\bar{t})^2}$ denote the between- and within-subject variance of the data and can be estimated by fitting a linear mixed effects model to the data. Here, Δ represents the treatment effect. We evaluated different values of Δ, when Δ=25% reflects a slowing of disease-related cognitive decline by at least 25%, attributed to the tested drug. Note that cognitive decline may be due to normal aging as well as AD-related pathology. Here, we remove the annualized cognitive decline due to normal aging so as not to overestimate the benefit of enrichment when computing the treatment effect.

We estimated and compared sample sizes for two groups of subjects. First, using data from all the mild AD subjects in the ADNI dataset that fit the selection criteria above (n=312); i.e., the unenriched group. Second, using only the subset of those ADNI MCI subjects identified as pMCI using baseline data and the classifier described above (n=64 for 2y); i.e., the enriched group.

## 3. Results

### 3.1. Prediction accuracy, sensitivity, and specificity

To assess how different features affect prediction accuracy, we trained models with different combinations of features. Table 2 shows the classification performance in terms of sensitivity, specificity, and accuracy, for all the models trained in this study, for two and three year follow up periods. Using hippocampal grading scores in addition to MoCA, ADAS13 and MMSE, yields the highest accuracy in predicting cognitive decline at 2 years. Comparing results between the classifier using only the baseline cognitive score and the corresponding classifier with the added MRI features showed that for both follow up periods, the accuracy of prediction is increased when adding MRI features.

**Table 2:** Classifier performances

| Feature sets (including Age) | 2 years follow-up | | | 3 years follow-up | | |
|---|---|---|---|---|---|---|
| | Sen (%) | Spec (%) | Acc (%) | Sen (%) | Spec (%) | Acc (%) |
| MoCA | 72.1±2.1 | 62.6±1.9 | 65.3±1.5 | 59.4±2.1 | 60.7±1.5 | 60.2±1.2 |
| ADAS13 | 71.2±2.5 | 71.3±1.6 | 71.3±1.3 | 67.4±0.9 | 68.3±2.0 | 68.8±1.4 |
| MoCA, ADAS13 | 74.8±2.4 | 74.7±1.2 | 74.7±1.1 | 66.4±1.8 | 70.6±1.7 | 70.4±1.3 |
| MoCA, ADAS13, MMSE | 76.5±1.5 | 75.7±1.3 | 75.9±1.0 | 65.2±1.9 | 70.4±1.5 | 70.8±1.2 |
| MoCA, ADAS13, MMSE, RAVLT | 76.1±2.1 | 74.8±1.2 | 75.2±0.9 | 66.3±1.8 | 69.3±1.6 | 71.0±1.3 |
| HC, EC | 76.2±2.1 | 70.1±1.3 | 71.7±1.1 | 75.1±1.9 | 68.9±1.4 | 71.0±1.3 |
| HC, ADAS13 | 78.8±1.3 | 73.6±1.2 | 74.7±1.2 | 75.7±1.6 | 70.7±1.4 | 72.6±1.1 |
| HC, MoCA | 75.9±2.4 | 72.4±1.4 | 73.2±1.2 | 71.6±2.0 | 67.3±1.2 | 69.0±0.9 |
| HC, EC, ADAS13 | 81.0±2.2 | 74.2±1.1 | 75.9±1.1 | 75.4±1.7 | 71.6±1.3 | 73.4±1.1 |
| HC, MoCA, ADAS13 | 80.4±1.6 | 74.3±1.1 | 75.8±0.9 | 75.4±2.2 | 70.7±1.2 | 72.8±1.1 |
| HC, EC, MoCA, ADAS13 | 80.2±2.1 | 75.0±0.8 | 76.7±0.7 | 74.9±2.4 | 73.0±1.4 | 74.0±1.2 |
| HC, MoCA, ADAS13, MMSE | 81.3±1.8 | 74.7±1.1 | 76.9±0.9 | 74.4±1.3 | 71.3±1.4 | 73.3±1.2 |
| HC, EC, MoCA, ADAS13, MMSE | 79.5±1.9 | 74.6±1.1 | 75.9±1.0 | 75.2±1.8 | 72.3±1.3 | 73.2±1.2 |

3.2 Power Analysis

Table 3 summarizes the CDR-SB values for the unenriched and enriched MCI cohorts that met the inclusion criteria described in section 2.1 and that were used to complete the power analysis.

| | Table 3. CDR-SB values. | | | |
|---|---|---|---|---|
| | Baseline mean (std dev) | Year 1 mean (std dev) | Year 2 mean (std dev) | Year 3 mean (std dev) |
| unenriched MCI | 1.631 (0.935) | 1.956 (1.457) | 2.356 (2.107) | 2.866 (2.978) |
| enriched MCI (2y) | 1.924 (0.875) | 2.68 (1.325) | 4.084 (2.504) | - |
| enriched MCI (3y) | 1.851 (0.874) | 2.559 (1.359) | 3.869 (2.639) | 5.094 (3.788) |

Figure 1 shows the required sample sizes for different treatment effects for both unenriched and enriched MCI cohorts. Using the unenriched group of MCI subjects, power analysis shows that 1075 subjects (764 subjects) per arm are required in a 2y (3y) trial of therapy with a hypothesized 25% effect size (80% power and 5% significance level) to reduce cognitive decline, measured by a two-point increase in CDR-SB (dotted lines in Fig. 1). When using the enriched cohort of MCI subjects, only 279 (273) subjects per arm are require for a 2-year (3-year) trial (solid lines in Fig. 3). These results demonstrate that enrichment using baseline HC, MoCA, ADAS13 and MMSE yields a 3.8-fold decrease in the sample size for a 2-year study (2.8-fold decrease for a 3-year study).

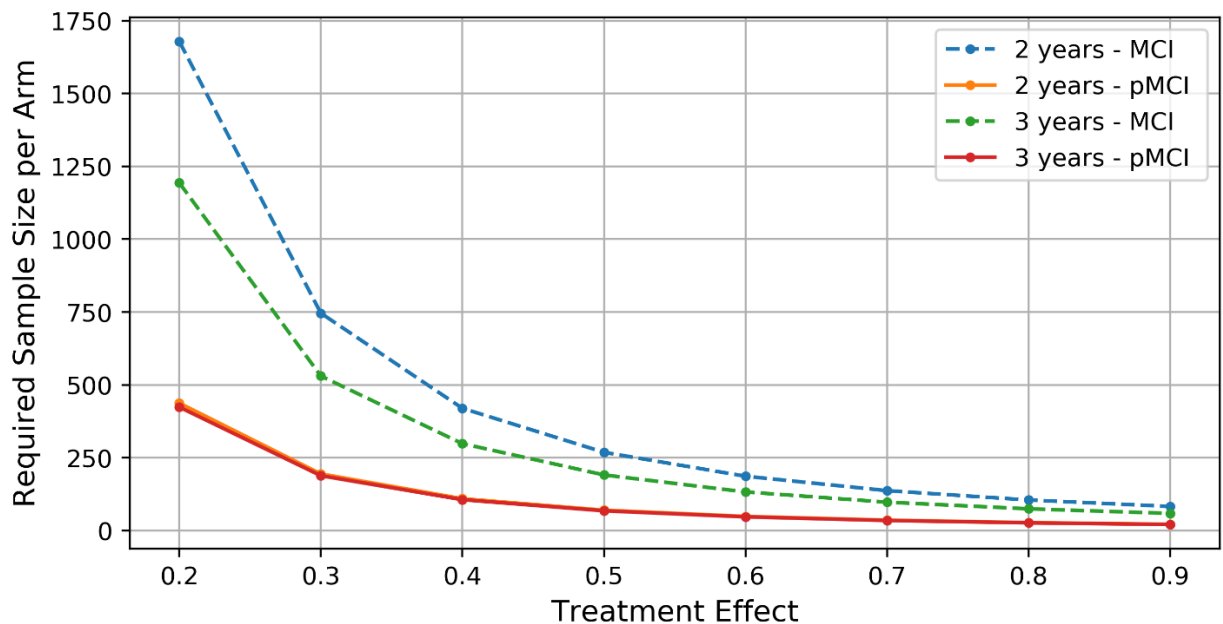

Figure 1: The required sample size per arm for different treatment effects.

## 4. Discussion

In the present study, we trained models to predict cognitive decline in patients in the early stages of AD dementia. We used feature sets consisting of baseline measures of either cognitive test scores, MRI-based grading scores, or a combination of both features for follow-up periods of 2 and 3 years in the ADNI dataset. The results demonstrate that cognitive test scores and our MRI-based features contribute differently to the result of the prediction, and combining cognitive test scores and MRI-based features improve prediction accuracy (Table 2). Using HC, MoCA, ADAS13 and MMSE as features yielded the highest prediction accuracy of 76.9% with a sensitivity of 81.3% and a specificity of 74.7% at 2 years.

In our previous work, we showed that when predicting onset of dementia in subjects with mild cognitive impairment, MRI-based features (SNIPE) are more sensitive compared to cognitive features, and even more so with longer follow-up periods, while cognitive features contribute more to the specificity of the prediction [15]. Here, we also show that cognitive features lose sensitivity when it comes to predicting cognitive decline at 36 months compared to that at 24 months.

Using MoCA and ADAS13 as features for our model, we achieved nearly 75% sensitivity for two-year prediction. By adding the HC SNIPE score to this feature set we were able to increase the sensitivity by 5.4% to 80.4%. At three years, prediction sensitivity of MOCA and ADAS dropped to 66.4%, but adding HC SNIPE features raises it to 75.4%, a 9% increase. As we have previously shown [15], MRI-driven features help contribute more sensitivity to the prediction at later follow-up periods. Despite the fact that predicting subtle cognitive decline is harder than predicting conversion from MCI to AD, the predictive accuracy of cognitive decline remains high.

We further estimated the statistical power of our prognostic model in terms of the sample size required to detect a treatment effect on the decline of cognitive abilities. The results of the current study showed that using our proposed prognostic model reduces the required sample size 3.8-fold in the MCI population to detect a 25% treatment effect reducing cognitive decline. This could give a marked clinical advantage, making the enrichment of the target cohort more precise with a smaller sample size, and therefore less costly.

Our results compare favorably to previous work. Lorenzi et al [30] evaluated a number of biomarkers to screen in subjects more likely to have cognitive decline. Without enrichment, their simulations required a sample size of 674 MCI patients per arm to detect a 25% treatment effect (90% power) on cognitive decline measured with CDR-SB in a two-year trial. Enrichment using either ADAS-COG, CSF tau, CSF Abeta42, CSF tau/abeta42, hippocampal volume, CSF p-tau, or [18f]-FDG PET decreased the number of patients required to 270, 310, 291, 264, 191, 287 and 240, respectively. At 191 subjects per arm, hippocampal volume offered a 3.5-fold reduction in the number of subjects required in their study. For direct comparison (25% effect, 90% power, 2-year trial), baseline HC SNIPE, MoCA, ADAS13, and MMSE enables a 3.8-fold reduction (from 1439 subjects unenriched to 375 with our classification tool). Ithapu et al [31] used deep learning techniques to evaluate enrichment in a 2-year trial of cognitive decline. They found that 1586 subjects were required to detect a 25% effect (80% power, significance of 0.05) without enrichment and that only 281 subjects were required per arm using baseline [F-18] fluorodeoxyglucose positron emission tomography (PET), amyloid florbetapir PET, and structural magnetic resonance imaging. While these results are very similar to ours, we do not require PET scanning, which is costly as well as invasive, and therefore less practical.

Recent work by Wolz et al. [32] evaluated enrichment in clinical trials in MCI using markers of amyloid (PET imaging or CSF analysis of beta amyloid) and neurodegeneration (measured by hippocampal volume) for a 25% effect size to decrease the rate of cognitive decline measured with MMSE or ADAS-Cog13 (with 80% power and significance level 0.05). While 908 unenriched subjects per arm were required for the ADAS-Cog13 outcome measure, this number could be reduced to 605 using baseline hippocampal volume, to 458 using baseline measures of amyloid, and 363 (corresponding to a 2.5-fold reduction) when using both. In a previous study, we have shown that SNIPE scores are better predictors of cognitive decline compared to volumetric measurements [33], and as a result, this score would further decrease the number of subjects needed, when used instead of volumetric-based measurements.

It is important to note that patient selection in clinical trials is an expensive process. At the beginning of a trial, one must screen a large number of subjects to select that meet eligibility criteria. This process currently involves the collection of several biomarkers (structural MRI, CSF biomarkers, amyloid/tau PET), but generally do not include prediction measures to identify subjects that are likely to have cognitive and functional decline. In ADNI, 35% of the MCI subjects in the study showed decline (at least two-point increase in CDR-SB) after 2 years. This shows the need to include roughly 3x more MCI subjects. With 58% of subjects declining after 3 years in ADNI, studies need to include almost twice as many subjects for this period. Decreasing the required sample size to demonstrate a clinical for effect would lead to massive savings in the follow-up visits of enrolled patients. Using a method to enrich the cohorts and decrease the number of subjects needed for a trial would therefore have a significance impact on the budget needed for such studies.

Our study is not without limitations. First of all, we did not use an independent dataset for testing our models. To further evaluate the accuracy of our study, independent validation using other datasets are needed. Additionally, the proportion of converters enrolled in the ADNI may change as MCI patients are followed for longer periods. Finally, here we measured against the decline in the unenriched MCI cohort from ADNI with specific inclusion criteria which might not necessarily be representative of real population of MCI subjects seen in trials or in the clinic.

## 5. Conclusion

In this work, we were able to predict future cognitive and functional decline in the early stages of AD using a prognostic model that combines cognitive scores and MRI-based biomarkers from a single baseline visit. These features are easy to measure, making this method efficient for clinicians to use as an aid to guide psycho-social interventions for individual patients based on their individual short-term prognosis. Refining clinical trial cohorts to the enrollment of subjects in the early stages of AD with a higher chance of declining over a shorter period of time could improve the efficiency of these trials.

**Supplementary Information:**

To ensure that the imbalance in the data does not bias the results, we also reported the AUC for each of the models, and demonstrated the ROC curve for two models: One consisting of all cognitive features and the other using both cognitive and MRI features (Figure S1).

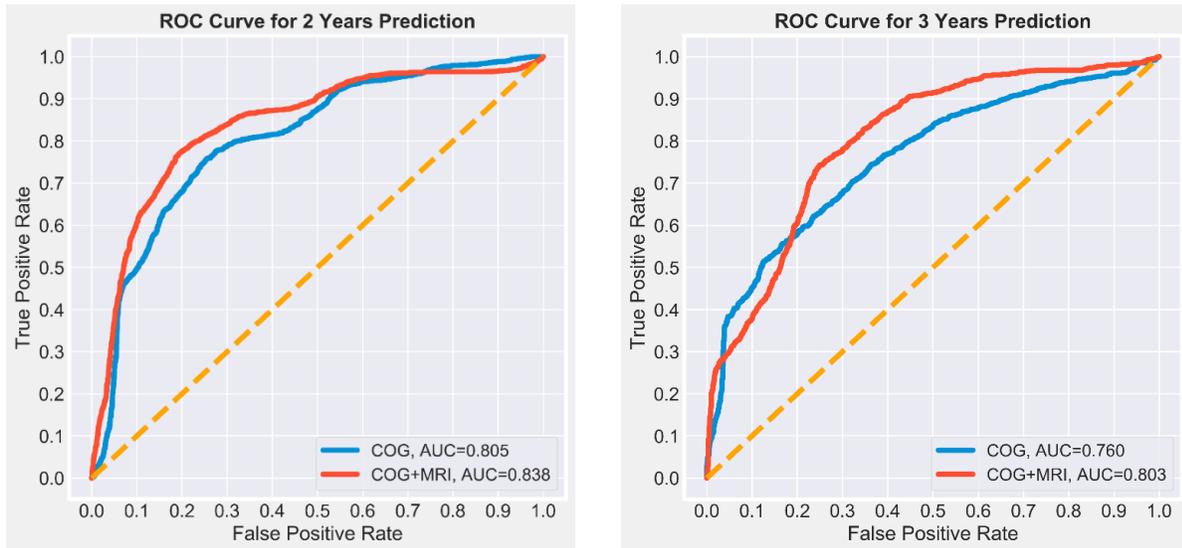

Figure S1: ROC curve of the model for both follow-up periods

Figure S2 shows the importance of each feature for both follow-up periods, showing how much each of these features affect prediction accuracy. To calculate the importance, we used the mean decrease of the impurity [34]. The impurity decrease for each of these features were averaged over all the nodes in all of the trees, using the Gini index as the impurity measure. In both models, ADAS is the most important feature followed by SNIPE scores for HC and EC. At three years, both ADAS and HC SNIPE gain importance while the two other cognitive scores fall behind. The increase in the importance of the SNIPE scores in time, shows its ability to detect AD-related changes earlier than most cognitive scores. The lowest importance was achieved for age and sex comparing to other features. We further tested for the significance difference of the feature importance between the two follow-up periods. The correction for multiple comparison showed the difference for all features is significant, except for EC and RAVLT.

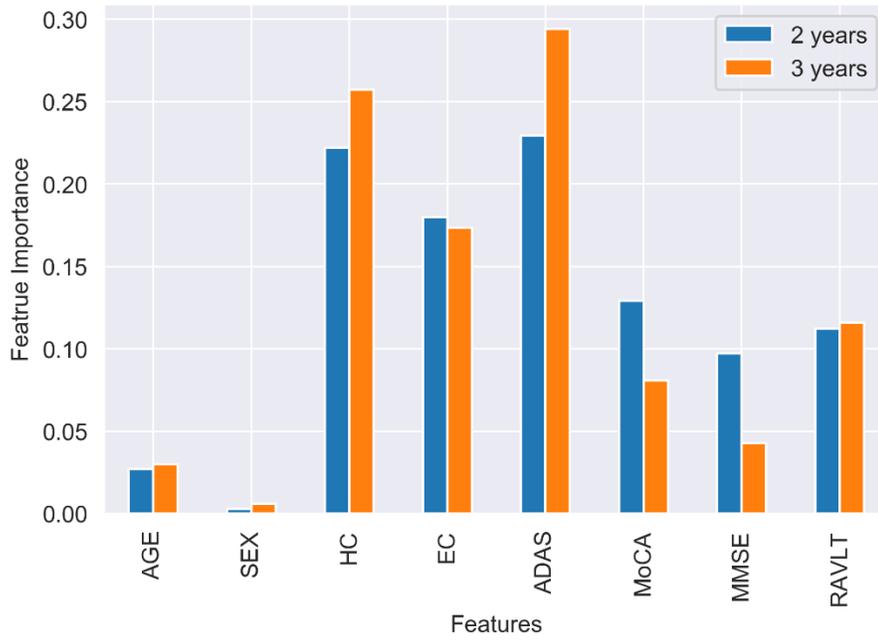

Figure S2: Feature importance for both follow-up periods